\documentstyle [multicol,aps,prb]{revtex}
\tightenlines
\begin{document}                                                    
\draft                                                      
\title {Screened interaction and self-energy in an infinitesimally     
polarized electron gas via the Kukkonen-Overhauser method } 
\author{Sudhakar Yarlagadda} 
\address{Saha Institute of Nuclear Physics, 1/AF Bidhannagar, 
Calcutta, India \\
 and Purdue University, W. Lafayette, Indiana, U.S.A. }
\author{ Gabriele F. Giuliani}
\address{Purdue University, W. Lafayette, Indiana, U.S.A. }
\date{\today}
\maketitle
\begin{abstract}
The screened electron-electron interaction 
$W_{\sigma, \sigma ^{\prime}}$ and the electron self-energy in
an infinitesimally polarized electron gas are derived by extending 
the approach of Kukkonen and Overhauser.
Various quantities in the expression for $W_{\sigma , \sigma ^{\prime}}$
are identified in terms of the relevant response functions of the
electron gas. The self-energy is obtained from 
$W_{\sigma , \sigma ^{\prime}}$ by making use of the GW method which 
in this case represents a consistent approximation. 
Contact with previous calculations is made.
\end{abstract}

                                                            

\section{Introduction}
Kukkonen and Overhauser \cite{KO} (KO) proposed an approximate analytic 
scheme for calculating the effects of exchange and correlations 
in an electron gas which accounts for both charge and spin 
fluctuations. The main merits of the KO method are its simplicity 
and physical clarity. One of the main results of the KO theory was
an expression for the quasiparticle effective interaction for a
paramagnetic electron gas.
Although this was not initially appreciated, these results 
were later confirmed for the paramagnetic case, by means of a more 
complex, less physically transparent, diagrammatic technique 
by Vignale and Singwi \cite{VS}. The diagrammatic analysis was 
then extended to the case of an infinitesimally polarized electron 
gas by Ng and Singwi \cite{NS}.
The situation was eventually clarified by the present authors 
who derived equivalent results within the framework of a theory 
of the electron gas based on the concept of quasiparticle pseudo 
hamiltonian 
\cite{YGSSC,YGpseudo}. 
This theory found successful application to the study of 
many-body effects in two-dimensional electronic 
systems 
\cite{YG2dliquids}. 

Even in the simple scenario of the electron gas, electronic 
correlations can be satisfactorily handled by analytic means
only in the high density regime where the random-phase-approximation 
\cite{gb,PN} provides a rigorous approach.
Earlier attempts to go beyond this simple scheme at metallic 
densities involved explicitly including short range exchange 
and correlation effects corrections in the charge response 
\cite{HUB,STLS}. 
In particular, the original contribution by Hubbard, i.e. the 
introduction of the so called many-body local field \cite{HUB}, 
established a useful physical framework within which the 
description of what amounts in practice to vertex corrections, 
became possible.  
KO were the first to exploit this methodology to its fullest extent
for the case of paramagnetic jellium.

A popular alternative approach for calculating the physical properties 
of the Landau quasiparticles in an electron gas is represented by 
the total energy method\cite{Rice}. In this approach a key step is 
represented by the determination of a suitable expression for 
the electron gas total energy as a functional of the particle 
occupation numbers. 
Although the procedure is quite standard and has been in use for quite 
some time, it was only recently realized that, in order to be able 
to achieve a correct microscopic theory, it is necessary to carefully 
keep separate track of the spin up and spin down occupation 
numbers \cite{YGcoupling,YGpseudo}. 
Accordingly even when studying the physics of an electron gas in 
its paramagnetic state, it is necessary, within this framework, 
to determine the energy of an infinitesimally polarized electron 
gas. This problem was tackled in 
Ref.\ \onlinecite{YGpseudo} via the pseudo 
hamiltonian method.
The self-energy obtained by this procedure proved to be equivalent 
to that independently derived by Ng and Singwi \cite{NS}.

The purpose of the present paper is to generalize the simple, elegant 
procedure developed by KO to the case of an infinitesimally polarized 
electron gas. To obtain a result useful also for multi-component
systems, in Sec. II we derive the screened
interaction between two electrons
by generalizing the theories of Refs.\ \onlinecite{KO} 
and \onlinecite{HEDLUN} 
to an infinitesimally polarized degenerate multi-valley system. 
In Sec. III the electron self-energy is obtained in a consistent 
fashion by making use of the lowest order diagram within what is 
commonly referred to as the GW approximation \cite{HEDIN}.  
We also show that the self-energy obtained following this 
procedure, although not identical, is very similar to that 
derived by the present authors in Ref.\ \onlinecite{YGpseudo}.
Lastly, in Sec. IV we present our conclusions.

\section{Effective interaction}
The first step in the KO procedure consists in obtaining a suitable
expression for the total effective potential felt by any given electron
of the liquid as a result of the introduction of a perturbing electron.
To this purpose we introduce a spin up electron, represented 
by a (number) density of Fourier amplitude $\rho _{\uparrow}$,
into the Fermi sea. Let $\Delta n _{\sigma}$ be the linear
density fluctuation of spin $\sigma = \pm 1$, set up by the 
introduction of this electron, and let 
$G^{\sigma , \sigma ^{\prime}}_{x(c), intra (inter)}$ 
be the appropriate generalized many-body local fields. 
Here the subscripts $x$ and $c$ refer respectively to exchange and 
correlation, while the labels $intra$ and $inter$ refer respectively 
to intra-valley and inter-valley processes.
Then, on assuming that the density fluctuations in all the valleys
are the same, and by following the standard linear response analysis, 
a complete expression for the potential felt by a spectator electron 
of opposite spin (down in this case) in the Fermi sea can be 
written as:
\begin{eqnarray}
\phi_{\downarrow \uparrow }= v(q) &&
\left \{ \left[\rho_{\uparrow }
+ \Delta n_{\uparrow} + \Delta n_{\downarrow} \right ]
- \left [ G^{\downarrow, \downarrow}_{x, intra} 
+ G^{\downarrow, \downarrow}_{c, intra} 
\right .  \right . 
\nonumber  \\
&& \left . 
+ (\nu _{v} -1) 
 G^{\downarrow, \downarrow}_{c, inter}
\right ]
\frac{ 2 \Delta n _{\downarrow}}{\nu_{v}} 
\nonumber \\
&& \left .  - \left [ G^{\downarrow , \uparrow}_{c, intra} +
 G^{\downarrow , \uparrow}_{c, inter}(\nu _{v} -1) \right ]
\frac{ 2 \rho_{\uparrow}+2 \Delta n _{\uparrow}}{\nu_{v}} \right \} ,
\label{Wdownarrow}
\end{eqnarray}
where it is understood that the potential
$\phi_{\uparrow \uparrow}$,
the density fluctuations, and the many-body local fields 
$G$ are all functions of both $\vec{q}$ and $\omega$. 
We immediately notice that, while the first term in this expression
represents the Hartree term, the remaining contributions stem from
exchange and correlation effects.
   
The potential felt by a spin up electron is obtained in a similar way.
One finds:
\begin{eqnarray}
\phi_{\uparrow \uparrow}= v(q) && \left \{ \left[\rho_{\uparrow }
+ \Delta n_{\uparrow} + \Delta n_{\downarrow} \right ]
- \left [ G^{\uparrow, \uparrow}_{x, intra} +
 G^{\uparrow, \uparrow}_{c, intra} 
\right . \right . 
\nonumber \\
&& \left .
+ (\nu _{v} -1) 
 G^{\uparrow, \uparrow}_{c, inter}
\right ]
\frac{2 \rho_{\uparrow} + 2 \Delta n _{\uparrow}}{\nu_{v}}
\nonumber \\
&&  \left . - \left [ G^{\uparrow, \downarrow}_{c, intra} +
 G^{\uparrow, \downarrow}_{c, inter}(\nu _{v} -1) \right ]
\frac{ 2 \Delta n _{\downarrow}}{\nu_{v}} \right \} ,
\label{Wuparrow}
\end{eqnarray}
where, in this case, explicit account has been taken of the extra 
exchange contribution arising from the fact that the perturbing and
the spectator electrons have the same spin.

These expressions can be simplified as follows.
As in Ref.\ \onlinecite{YGpseudo}, for the infinitesimally 
polarized case one can
assume the following relations: 
\begin{equation}
G^{\uparrow, \uparrow}_{x, intra} = 
G^{\downarrow, \downarrow}_{x, intra} ,
\label{Gx}
\end{equation}
and
\begin{equation}
  G^{\uparrow, \uparrow (\downarrow )}_{c, intra(inter)} =
 G^{\downarrow, \downarrow (\uparrow )}_{c, intra (inter)} ,
\label{Gc}
\end{equation}
which, strictly speaking, are valid (by symmetry) for an unpolarized 
electron gas.
Moreover, if upon scattering the electrons retain their valleys 
we also have:
\begin{equation}
  G^{\uparrow, \uparrow }_{c, inter} =
  G^{\uparrow, \downarrow }_{c, inter} =
  G^{\uparrow, \downarrow }_{c, intra} .
\end{equation}
Then, on defining the single valley local fields $G_{\pm}$ as
\begin{equation}
G_{\pm} \equiv
  G^{\uparrow, \uparrow }_{x, intra} +
  G^{\uparrow, \uparrow }_{c, intra} \pm
  G^{\uparrow, \downarrow }_{c, intra} ,
\end{equation}
and the multi-valley local fields  $G_{\pm}^{v}$ as
\begin{equation}
G_{+(-)}^{v} \equiv
G_{+}-G_{-(+)} + \frac{G_- }{\nu_v} , 
\end{equation}
the potential felt by an electron with spin $\sigma$ 
can be cast in the following compact form:
\begin{eqnarray}
\phi_{\sigma \uparrow} =  v(q) \left \{
(1- G^{v}_{+} - \right . & & \sigma G_{-}^{v}) \rho _{\uparrow} 
+  \left[
 \Delta n_{\uparrow} + \Delta n_{\downarrow} \right] (1- G_{+}^{v})
 \nonumber \\
 && \left .
- \sigma 
 \left[
 \Delta n_{\uparrow} - \Delta n_{\downarrow} \right ] G_{-}^{v}
 \right \} .
\label{Wsigma}
\end{eqnarray}
We next recognize that within linear response one can write: 
\begin{equation}
\Delta n _{\sigma} = \nu_{v} \chi_{0}^{\sigma}
\phi_{\sigma \uparrow} ,
\label{linresp}
\end{equation}
where $\chi_{0}^{\sigma}$ is the spin $\sigma$ response for a non 
interacting electron gas and can be expressed as follows:
\begin{equation}
\chi_{0}^{\sigma}( \vec{q} , \omega ) \equiv
\sum _{\vec{p}} \int _{-\infty}^{\infty} \frac {d \epsilon}{2 \pi i}
g^{\sigma} ( \vec{p} , \epsilon )
g^{\sigma} ( \vec{p} + \vec{q} , \epsilon + \omega ) .
\end{equation}
In this expression $g^{\sigma} ( \vec{p} , \omega )$
is the bare one electron Green's function given by
\begin{equation}
g^{\sigma} ( \vec{p} , \omega ) 
\equiv 
\frac{n^{\sigma}_{\vec{p}}}{\omega - \epsilon _{\vec{p}} - i \eta }
+ \frac{1 - n^{\sigma}_{\vec{p}}}{\omega - 
\epsilon _{\vec{p}} + i \eta } ,
\end{equation}
with $ n^{\sigma}_{\vec{p}}$ being the exact occupation number.
Then using Eqs.\ (\ref{Wsigma}) and (\ref{linresp}), we obtain the 
following relationships for the potentials: 
\begin{equation}
\phi_{\uparrow \uparrow}=\frac{v(q) \left [
(1- G^{v}_{+} -  G_{-}^{v}) 
+4 v(q) \nu _{v} \chi_{0}^{\downarrow}
 G_{-}^{v}(1- G_{+}^{v}) \right ]}{{\cal{D}}^{v}} \rho_{\uparrow} ,
\label{Wrelation1}
\end{equation}
and
\begin{equation}
\phi_{\downarrow \uparrow}=\frac{v(q) 
(1- G^{v}_{+} +  G_{-}^{v}) 
  }{{\cal{D}}^{v}} \rho_{\uparrow} ,
\label{Wrelation2}
\end{equation}
with ${\cal{D}}^{v}$ defined as follows:
\begin{eqnarray}
{\cal{D}}^{v} \equiv 
&& 1-v(q) 
\left ( \nu _{v} \chi_{0}^{\uparrow}+
\nu _{v} \chi_{0}^{\downarrow} \right )
\left ( 1- G^{v}_{+} -  G_{-}^{v} \right ) 
\nonumber \\
&&
-4 
v^2(q) 
\nu _{v}^2 
\chi_{0}^{\uparrow}
\chi_{0}^{\downarrow}
 G_{-}^{v}(1- G_{+}^{v}) .
\label{Dv}
\end{eqnarray}

At this point, in order to obtain the screened electron-electron 
interaction from the effective potentials $\phi_{\sigma \uparrow}$ 
in the KO method one argues as follows.
To correctly describe the physics of the problem, several different 
contributions stemming from exchange and correlation effects have 
been approximately accounted for through the local fields 
$G_{\pm}^{v}$ in the formulas for 
$\phi_{\sigma \uparrow}$. 
A physically satisfactory expression for the electron-electron 
screened interaction $W_{\sigma \uparrow}$ between two electrons can 
then be obtained by simply subtracting from such expressions the terms 
accounting for the explicit exchange and correlation contributions 
between the spectator and the perturbing electron. Accordingly 
following KO we write:
\begin{equation}
W_{\sigma \uparrow} ~ \rho _{\uparrow} =
\phi_{\sigma \uparrow} +v(q)
 \left [ ( G_{+}^{v} + \sigma G_{-}^{v})
\rho _{\uparrow} \right ] .
\label{effW}
\end{equation}
More generally, based on the isotropy of the unpolarized system, we have
for the spin dependent screened interaction  potential:
\begin{equation}
W_{\vec{\sigma _{1}} \vec{\sigma _{2}}}=
\frac{
W_{\uparrow \uparrow}  
+W_{\downarrow \uparrow}  }{2}
+ \vec{\sigma _{1}} \cdot \vec{\sigma _{2}}
\frac{
W_{\uparrow \uparrow}  
-W_{\downarrow \uparrow}  }{2} .
\label{Wisotropic}
\end{equation}
It is crucial to appreciate here that, although the exchange and 
correlation contributions to the effective potential between 
the spectator and the perturbing electron
have been explicitly removed, the resulting 
scattering matrix elements $M_{\alpha , \beta}$
between two antisymmetrized states of the interaction potential
$W_{\vec{\sigma _{1}} \vec{\sigma _{2}}}$ will automatically account 
for exchange and (to some extent) correlation effects \cite{ZO}.
This can be seen from:
\begin{eqnarray}
\label{matrixelem}
M_{\alpha , \beta} &
 = &
\frac{1}{2} 
\langle \psi_{f} |
W_{\vec{\sigma _{1}} \vec{\sigma _{2}}}(\vec{r}_{1}-
\vec{r}_{2}, \omega )
 | \psi_{i} \rangle 
\nonumber \\
 &
= &
W_{\alpha \beta}(\vec{q} , \omega ) -  
\delta_{\alpha , \beta }
W_{\alpha \alpha}(\vec{k}_{1}- \vec{k}_{2} -\vec{q} , \omega ) 
\nonumber \\
&&
- \delta_{-\alpha , \beta }
W^{T}(\vec{k}_{1}- \vec{k}_{2} -\vec{q} , \omega ) , 
\label{Malphabeta}
\end{eqnarray}
where, with obvious notation:
\begin{equation}
| \psi_{i} \rangle \equiv
|\vec{k}_{1},\alpha; \vec{k}_{2},\beta \rangle
- |\vec{k}_{2},\beta; \vec{k}_{1},\alpha \rangle ,
\end{equation}
\begin{equation}
\langle \psi_{f} | \equiv
\langle \vec{k}_{1}-\vec{q},\alpha; \vec{k}_{2}+\vec{q},\beta|
- \langle \vec{k}_{2}+\vec{q},\beta; \vec{k}_{1}-\vec{q},\alpha| ,
\end{equation}
we have defined
\begin{equation}
W^{T} (\vec{q} , \omega ) 
 =
W_{\uparrow \uparrow}
(\vec{q} , \omega ) 
 - W_{\downarrow \uparrow}
(\vec{q} , \omega ) .
\end{equation}
In Eq.\ (\ref{matrixelem}),
$W_{\vec{\sigma _{1}} \vec{\sigma _{2}}}(\vec{r}, \omega )$ is the 
real space Fourier transform of the screened interaction
given in Eq.\ (\ref{Wisotropic}). 
Thus the matrix elements $M_{\sigma \sigma}$ automatically incorporate
antisymmetrization effects and quite naturally contain non-local
contributions.

As for $W_{-\sigma \sigma}$, the correlation
effects between two opposite 
spin Fermi sea electrons are accounted through spin-flip scattering 
processes with the corresponding screened potential in the transverse
(spin-flip) channel being given by
\begin{equation}
W^{T} =
-2 \mu_{B}^{-2} \left [ v(q)  G_{-}^{v}
(\vec{q} , \omega ) \right ]^{2} \chi_S
(\vec{q} , \omega ) ,
\label{Woppspin}
\end{equation}
which is twice the contribution from the longitudinal 
spin fluctuations.

Now, for an infinitesimally  polarized system, while the matrix elements
$M_{\alpha , \beta} $ are still given by Eq.\ (\ref{Malphabeta}), for
$W^{T}$ we propose the following natural ansatz based on the structure
of its unpolarized counterpart:
\begin{eqnarray}
W^{T}  && =
\frac{
W_{\uparrow \uparrow}
(\vec{q} , \omega ) 
+W_{\downarrow \downarrow}
(\vec{q} , \omega ) 
}{2} -
W_{\downarrow \uparrow}
(\vec{q} , \omega ) 
\nonumber \\
&& =
-2 \mu_{B}^{-2} \left [ v(q)  G_{-}^{v}
(\vec{q} , \omega ) \right ]^{2} \chi_S
(\vec{q} , \omega ) .
\label{WT}
\end{eqnarray}

Using arguments similar to those presented above
to obtain $\phi_{\sigma \uparrow}$, the charge
response $\chi_{C}$, the spin response $\chi_{S}$,
 and the mixed charge-spin
responses $\chi_{CS}^{V}$
$(\equiv \frac {\Delta n_{\uparrow}
-\Delta n_{\downarrow}}{\phi_{ext}}) $ and
 $\chi_{CS}^{H}$
$(\equiv
\frac{ \Delta n_{\uparrow}
+\Delta n_{\downarrow}}{\mu_{B}H^{z}_{ext}})$
 can be obtained to be (see Ref.\ \onlinecite{YGpseudo} for
details):
\begin{equation}
\chi_{C} \equiv \frac{
\Delta n_{\uparrow}
+\Delta n_{\downarrow}}{\phi_{ext}} =
\frac{  \nu _{v} \chi_{0}^{\uparrow}+
\nu _{v} \chi_{0}^{\downarrow}
+4 v(q) \nu _{v}^2 
\chi_{0}^{\uparrow}
\chi_{0}^{\downarrow}
 G_{-}^{v}}{{\cal{D}}^{v}} ,
\label{chiC}
\end{equation}
\begin{eqnarray}
\chi_{S} && \equiv  \mu_{B} \frac{
\Delta n_{\downarrow}
-\Delta n_{\uparrow}}{H^{z}_{ext}} 
\nonumber \\
&& = -\mu_{B}^2
\frac{  \nu _{v} \chi_{0}^{\uparrow}+
\nu _{v} \chi_{0}^{\downarrow}
-4 v(q) \nu _{v}^2 
\chi_{0}^{\uparrow}
\chi_{0}^{\downarrow}
(1- G_{+}^{v})}{{\cal{D}}^{v}} ,
\label{chiS}
\end{eqnarray}
and
\begin{equation}
\chi_{CS}^V  = 
\chi_{CS}^H
=
\frac{ ( \nu _{v} \chi_{0}^{\uparrow}-
\nu _{v} \chi_{0}^{\downarrow})}
{{\cal{D}}^{v}} .
\label{chiCS}
\end{equation}
The transverse spin response $\chi^{T\sigma}$ can be defined
for a multi-valley system
as follows:
\begin{equation}
\chi^{T\sigma} \equiv
- \mu_{B}^{2} \frac{\nu_{v} {\chi_0}^{T\sigma}}{
1+2v(q) G_{-}^{Tv}\nu_{v}\chi_{0}^{T\sigma}} ,
\label{transresp}
\end{equation}
where
\begin{equation}
\chi_{0}^{T\sigma}( \vec{q} , \omega ) \equiv
\sum _{\vec{p}} \int _{-\infty}^{\infty} \frac {d \epsilon}{2 \pi i}
g^{-\sigma} ( \vec{p} , \epsilon )
g^{\sigma} ( \vec{p} + \vec{q} , \epsilon + \omega ) .
\end{equation}
The only undefined term in Eq.\ (\ref{transresp}) 
is the transverse many-body
local field $G_{-}^{Tv}$ for which we will assume the relationship:
\begin{eqnarray}
G_{-}^{Tv} = \frac{1}{2\nu_v}
\left [
G_{x,intra}^{\uparrow \uparrow}
\right . &&
+G_{c,intra}^{\uparrow \uparrow}
+G_{x,intra}^{\downarrow \downarrow}
\nonumber \\
&& 
\left . 
+G_{c,intra}^{\downarrow \downarrow}
-2G_{c,intra}^{\uparrow \downarrow} \right ] .
\end{eqnarray}
Here it must be pointed out that with the approximations made
 in Eqs.\ (\ref{Gx}) and (\ref{Gc}),
$G_{-}^{Tv}$ coincides with the longitudinal field $G_{-}^{v}$.

Using Eqs.\ (\ref{Wrelation1})-(\ref{effW}) and
Eqs.\ (\ref{chiC})-(\ref{chiCS})
it can finally be shown that
\begin{eqnarray}
W_{\sigma \sigma}
(\vec{q} , \omega )
= &&
v(q) \left \{1+v(q) \left [ 1 - G_{+}^{v}
(\vec{q} , \omega ) \right ]^{2} \chi_C
(\vec{q} , \omega ) \right \}
\nonumber \\
&& - \mu_{B}^{-2} \left [ v(q)  G_{-}^{v}
(\vec{q} , \omega ) \right ]^{2} \chi_S
(\vec{q} , \omega ) 
\nonumber \\
&& - 2\sigma v(q)^2
  G_{-}^{v}
(\vec{q} , \omega ) 
\left [1 -   G_{+}^{v}
(\vec{q} , \omega ) \right ] \chi_{CS}
(\vec{q} , \omega ) ,
\label{Wsigmasigma}
\end{eqnarray}
and that 
\begin{eqnarray}
W_{\downarrow \uparrow}
(\vec{q} , \omega )
= &&
v(q) \left \{1+v(q) \left [ 1 - G_{+}^{v}
(\vec{q} , \omega ) \right ]^{2} \chi_C
(\vec{q} , \omega ) \right \}
\nonumber \\
&& + \mu_{B}^{-2} \left [ v(q)  G_{-}^{v}
(\vec{q} , \omega ) \right ]^{2} \chi_S
(\vec{q} , \omega ) .
\label{Wdownup}
\end{eqnarray}
\section{Self-energy}
The screened interaction
$W_{\sigma \sigma ^{\prime}} (\vec{q} , \omega )$
given in Eqs.\ (\ref{Wsigmasigma}) and (\ref{Wdownup})  is similar
 to the effective screened interaction
$V_{\sigma , \sigma^{\prime}} ( \vec{q}, \omega, \omega, \omega )$
derived by the present authors 
(see Eq.\ (31) of Ref.\ \onlinecite{YGpseudo}). In fact, if 
in $V_{\sigma , \sigma^{\prime}} ( \vec{q}, \omega, \omega, \omega )$
the real response functions are replaced by the full complex responses
and the complex conjugate many-body local fields that are
pre-factors to the response functions are replaced by their
complex counterparts, one gets
exactly $W_{\sigma \sigma ^{\prime}} (\vec{q} , \omega ) -v(q)$.
As argued in Ref.\ \onlinecite{YGpseudo}, the effective screened 
interaction 
$V_{\sigma , \sigma^{\prime}} ( \vec{q}, \omega, \omega, \omega )$
should be used for calculations carried out up to first order only.
This conclusion is supported by the results of the elegant analysis
carried out by Takada in Ref.\ \onlinecite{Takada}. 
To evaluate higher order terms would in this case not only not lead to 
better results but would in fact be erroneous.
It is then quite reasonable to evaluate the quasiparticle self-energy 
to first order in the screened interaction from the expression:
\begin{eqnarray}
\Sigma^{\sigma}( \vec{p} , \omega ) =
-\sum_{\vec{q}} \int _{-\infty}^{\infty} \frac {d \epsilon}{2 \pi i}
\left \{  \right . && W_{\sigma \sigma}
g^{\sigma} ( \vec{p} - \vec{q} , \omega - \epsilon  ) 
\nonumber \\
&& \left .
+ W_{ \sigma}^{T}
g^{-\sigma} ( \vec{p} - \vec{q} , \omega - \epsilon  ) \right \} ,
\label{selfeng}
\end{eqnarray}
where
$ W_{\sigma \sigma}$
 is given by
 Eq.\ (\ref{Wsigmasigma})
and $W_{ \sigma} ^{T}$ is defined as follows:
\begin{equation}
W_{ \sigma} ^{T}
(\vec{q} , \omega )
\equiv 
-4 \mu_{B}^{-2} \left [ v(q)  G_{-}^{Tv}
(\vec{q} , \omega ) \right ]^{2} \chi^{T\sigma}
(\vec{q} , \omega ) .
\label{WTsigma}
\end{equation}
In the above equation for
$\Sigma^{\sigma}( \vec{p} , \omega ) $ it is understood that
 $ W_{\sigma \sigma}$ and $W_{ \sigma} ^{T}$ are defined
in terms of time ordered response functions
 and many-body local fields.
 Furthermore, the above expression for $W_{ \sigma} ^{T}$ 
has been obtained from
Eq.\ (\ref{Woppspin})
after noting that in the transverse channel we expect the
screened interaction potential to be determined by the transverse 
spin susceptibility. 

Earlier on, in  Ref.\ \onlinecite{YGpseudo},  the present authors
derived the following  expression for the self-energy 
of an infinitesimally polarized Fermi gas:
\begin{eqnarray}
\Sigma^{\sigma}( \vec{p} , \epsilon_{\vec{p}}^{\sigma} ) =
&& -\sum_{\vec{q}}
\left \{  n^{\sigma}_{\vec{p} - \vec{q}} 
Re \left [ v(q) +D_{1} ( \vec{q} , 
 \epsilon_{\vec{p}}^{\sigma}
 -\epsilon_{\vec{p}-\vec{q}}^{\sigma} ) \right ]
\right . 
\nonumber \\
&& \left .
+  n^{-\sigma}_{\vec{p} - \vec{q}} Re \left [ D_{2} ( \vec{q} , 
 \epsilon_{\vec{p}}^{\sigma}
 -\epsilon_{\vec{p}-\vec{q}}^{\sigma} ) \right ] \right .
\nonumber \\
&&
\left . ~~~
-P \int_{0}^{\infty} \frac{d \omega}{\pi} \left [
\frac{Im[D_{1}( \vec{q} , \omega ) ]}
 {\omega -\epsilon_{\vec{p}}^{\sigma}
+\epsilon_{\vec{p}-\vec{q}}^{\sigma} }
+ \frac{Im[D_{2}( \vec{q} , \omega ) ]}
 {\omega -\epsilon_{\vec{p}}^{\sigma}
+\epsilon_{\vec{p}-\vec{q}}^{-\sigma} }
 \right ] \right \} ,
\label{scexchole}
\end{eqnarray}
where
\begin{eqnarray}
D_{1}(\vec{q} , \epsilon ) \equiv 
&&
v(q)^2 \left [ \left | 1 - G_{+}^{v} \right |^2
\chi_C
(\vec{q} , \epsilon )
- \mu_{B}^{-2} \left |  G_{-}^{v} \right | ^2
 \chi_S
(\vec{q} , \epsilon ) \right .
\nonumber \\
&&
\left .
-2 \sigma  Re[ G_{-}^{v}
(1 -   G_{+}^{v\star})]
 \chi_{CS}
(\vec{q} , \epsilon ) \right ] ,
\end{eqnarray}
and
\begin{equation}
D_{2}(\vec{q} , \epsilon ) \equiv 
 -4 \mu_{B}^{-2} v(q)^2 \left |   G_{-}^{Tv} \right | ^{2}
 \chi^{T\sigma}
(\vec{q} , \epsilon ) ,
\end{equation}
with the local fields $G_{\pm}$ being functions of $\vec{q}$ and
 $ \epsilon^{\sigma}_{\vec{p}} 
 - \epsilon^{\sigma}_{\vec{p} -\vec{q}}$ while $G_{-}^{T}$
being a function of  
 $ \epsilon^{\sigma}_{\vec{p}} 
 - \epsilon^{-\sigma}_{\vec{p} -\vec{q}}$ .

The expression for the self-energy given in Eq.\ (\ref{scexchole})
can be rearranged, as will be shown below, to give
the following expression similar to that of 
Eq.\ (\ref{selfeng}) derived above:
\begin{eqnarray}
\Sigma^{\sigma}( \vec{p} , \omega ) =
-\sum_{\vec{q}} 
&&
\int _{-\infty}^{\infty} \frac {d \epsilon}{2 \pi i}
\left \{  \left [ v(q) +
D_{1}(\vec{q}, \epsilon ) \right ]
g^{\sigma} ( \vec{p} - \vec{q} , \omega - \epsilon  ) 
\right .
\nonumber \\
&& ~~~
\left .
+D_{2}(\vec{q}, \epsilon )
g^{-\sigma} ( \vec{p} - \vec{q} , \omega - \epsilon  ) \right \} .
\label{selfengold}
\end{eqnarray}
Now, if in  Eq.\ (\ref{selfengold}) the complex conjugate local fields 
are replaced by complex local fields and the frequencies of the local 
fields that are pre-factors to the response functions are replaced by 
those of the response functions, we then get exactly the self-energy
given by Eq.\ (\ref{selfeng}).
We further note that the expression for the self-energy as given by 
Eq.\ (\ref{selfeng}) is identical to the result of 
Ref.\ \onlinecite {NS}.

We will now rearrange the expression for the self-energy
given in Eq.\ (\ref{selfengold})
in terms of screened exchange and coulomb hole contributions.
It can be verified from Kramers-Kronig relations that
$D_{1,2}(\vec{q} , \epsilon )$ can be cast in the following form:
\begin{equation}
D_{1,2}(\vec{q} , \epsilon )
=-\int _{0}^{\infty} \frac{dt}{\pi}
\left \{
\frac{Im D_{1,2} (\vec{q} , t )}{\epsilon -t + i \eta }
-\frac{Im D_{1,2} (\vec{q} , -t )}{\epsilon +t - i \eta } \right \} .
\label{lehrep}
\end{equation}
Noting that 
$D_{1,2} (\vec{q} , \epsilon)$ vanishes for large values of $\epsilon$,
we readily obtain
\begin{eqnarray}
&& i\sum_{\vec{q}} \int _{-\infty}^{\infty} \frac {d \epsilon}{2 \pi }
  D_{1,2} ( \vec{q} , \epsilon )
g^{\sigma} ( \vec{p} - \vec{q} , \omega - \epsilon  )  
\nonumber \\
&&
=-\frac{i}{2\pi^2}\sum_{\vec{q}}(1-n^{\sigma}_{\vec{p}-\vec{q}})
\int_{0}^{\infty} dt [Im  D_{1,2} ( \vec{q} , t) ]
\int_{-\infty}^{\infty} d\epsilon
\frac{1}{
[\omega -\epsilon -\epsilon ^{\sigma}_{\vec{p}-\vec{q}} +i\eta]
[\epsilon -t +i\eta]}
\nonumber \\
&&
+\frac{i}{2\pi^2}\sum_{\vec{q}}n^{\sigma}_{\vec{p}-\vec{q}}
\int_{0}^{\infty} dt [Im D_{1,2} ( \vec{q} , -t) ]
\int_{-\infty}^{\infty} d\epsilon
\frac{1}{
[\omega -\epsilon -\epsilon ^{\sigma}_{\vec{p}-\vec{q}} -i\eta]
[\epsilon + t -i\eta]}
\nonumber \\
&&
=\sum_{\vec{q}}n^{\sigma}_{\vec{p}-\vec{q}}
\int_{0}^{\infty} \frac{dt}{\pi} \left \{
\frac{ Im D_{1,2} ( \vec{q} , t)}
{\omega -\epsilon ^{\sigma}_{\vec{p}-\vec{q}} -t+i\eta}
- \frac{ Im D_{1,2} ( \vec{q} , -t)}
{\omega -\epsilon ^{\sigma}_{\vec{p}-\vec{q}} +t-i\eta} \right \}
\nonumber \\
&&
-\sum_{\vec{q}}
\int_{0}^{\infty} \frac{dt}{\pi}
\frac{Im  D_{1,2}( \vec{q} , t)}
{\omega -\epsilon ^{\sigma}_{\vec{p}-\vec{q}} -t+i\eta}
\nonumber \\
&&
=-\sum_{\vec{q}}n^{\sigma}_{\vec{p}-\vec{q}}
 D_{1,2}
( \vec{q} , \omega - \epsilon ^{\sigma}_{\vec{p}-\vec{q}})
-\sum_{\vec{q}}
\int_{0}^{\infty} \frac{dt}{\pi}
\frac{ Im D_{1,2}
( \vec{q} , t)}
{\omega -\epsilon ^{\sigma}_{\vec{p}-\vec{q}} -t+i\eta} .
\label{D12manip}
\end{eqnarray}
 Finally by noting that
\begin{equation}
\sum_{\vec{q}} \int _{-\infty}^{\infty} \frac {d \epsilon}{2 \pi i}
v(q)
g^{\sigma} ( \vec{p} - \vec{q} , \omega - \epsilon  ) 
=\sum_{\vec{q}}v(q)n^{\sigma}_{\vec{p}-\vec{q}} ,
\label{hfeng}
\end{equation}
we see from Eqs.\ (\ref{D12manip}) and (\ref{hfeng}) that the 
self-energy given in Eq.\ (\ref{selfengold})
is equivalent  to the expression in Eq.\ (\ref{scexchole}).

\section{conclusions}
We have shown that the Kukkonen-Overhauser approach
to derivation of the screened interaction between two electrons in
an interacting electron liquid can be extended to the case of 
an infinitesimally polarized electron gas provided one makes a
reasonable ansatz for the spin-flip term.
The screened interaction obtained
in this approach can be then used to obtain the electron self-energy
by means of a GW type of approximation. The self-energy
thus obtained is similar to that previously derived by different means 
by the present authors.
\begin{center}
   {\bf ACKNOWLEDGMENTS}                                   
\end{center}
The authors acknowledge  many insightful conversations on the 
subject with A. W. Overhauser. 
GFG wishes to acknowledge the support of DOE Grant 
\# DE-FG02-90ER45427 through the Midwest Superconductivity Consortium.

%
\end{document}